\documentclass[
reprint,
superscriptaddress,
amsmath,amssymb,
prl,
]{revtex4-2}

\usepackage{graphicx}
\usepackage{dcolumn}
\usepackage{bm}
\usepackage{chemformula}
\usepackage{hyperref}
\usepackage{amsfonts,amssymb}
\usepackage[english]{datetime2}
\usepackage{tabularx}
\usepackage{amsmath}
\usepackage{float}
\usepackage{array}
\usepackage{booktabs}
\usepackage{graphicx}
\usepackage{makecell}
\newcommand{\Arg}{\mathrm{Arg}}

\usepackage[version=4]{mhchem}

\usepackage{physics} 

\begin{document}
	
	\title{M\"obius-topological auxiliary function for $f$ electrons}
	
	\author{Biaoyan~Hu}
	\email{hubiaoyan@quantumsc.cn}
	\affiliation{Quantum Science Center of Guangdong-Hong~Kong-Macao Greater Bay Area, Shenzhen 518045, China}
	\affiliation{Department of Physics, Southern University of Science and Technology, Shenzhen 518055, China}
	
	\begin{abstract}
		$f$-electron systems exhibit a subtle interplay between strong spin--orbit coupling and crystal-field effects, producing complex energy landscapes that are computationally demanding. We introduce auxiliary functions, constructed by extending hydrogen-like wave functions through a modification of the Legendre function. These functions often possess a M\"obius-like topology, satisfying $\psi(\varphi) = -\psi(\varphi + 2\pi)$, while their squared modulus respects inversion symmetry. By aligning $|\psi|^2$ with the symmetry of the crystal field, they allow rapid determination of eigenstate structures without the need for elaborate calculations. The agreement with established results indicates that these functions capture the essential physics while offering considerable computational simplification.
	\end{abstract}
	
	\maketitle
	
	\emph{Introduction.—}
	Understanding electronic eigenstates in solids is essential for interpreting their magnetic, spectroscopic, and thermodynamic properties. An early quantum--mechanical treatment of atomic term splitting in crystals was provided by Bethe \cite{bethe1929termaufspaltung}, who analyzed how crystalline electric fields lift electronic degeneracies. This laid the foundation for crystal field theory developed by Van Vleck \cite{Vanvleck1932}, providing a general framework applied to magnetic and spectroscopic phenomena. Since then, crystal field theory has guided the interpretation of spectra and thermodynamic behavior in minerals with transition metals \cite{burns1993mineralogical}, and clarified structure-property relations in luminescent materials \cite{song2022basic}. More recently, crystal-field analysis has become a key tool for probing rare-earth magnetism and correlated $f$-electron systems \cite{kabeya2022eigenstate, ueta2022anomalous, uzoh2023influence, bordelon2023structural, ram2024crystalline, guchhait2024magnetic, guchhait2025magnetic}.  
	
	Formal links between quantum operators and classical multipoles have been established \cite{kusunose2008description}. In particular, the crystal-field Hamiltonian can be expressed via the Stevens operator method \cite{stevens1952matrix} as $H_\mathrm{CF} = \sum_{l,m} B_l^m O_l^m$, where $B_l^m$ and $O_l^m$ are the crystal-field parameters and Stevens operators, respectively \cite{hutchings1964point}.
	
	Existing methods for $f$-electron systems are fully capable but often cumbersome due to proliferating notations and intricate derivations. A more direct approach yielding explicit eigenstate structures would greatly simplify analysis of these technically demanding systems.
	
	\emph{Auxiliary function.—}
	Motivated by the need for a simpler approach, we introduce a modified theoretical framework based on the hydrogen-like Schr\"odinger equation. By modifying the Legendre function, we construct functions for all angular momenta, including half-integer values, enabling direct determination of $f$-electron eigenstate structures.
	
	The Schr\"odinger equation for an electron under Coulomb potential in Gaussian units is given by: 
	\begin{equation}\label{Schrodinger}
		\left(-\frac{\hbar^2}{2m_{\rm e}}\nabla^2-\frac{Ze^2}{r}\right)\psi(r,\theta,\varphi)=E\psi(r,\theta,\varphi),
	\end{equation}
	where $m_{\rm e}$ is the electron mass and $\nabla^2$ is the Laplacian operator. 
	This equation is generally applicable only to hydrogen-like atoms, but with Slater's rules \cite{slater1930atomic}, it can be approximately applied to multi-electron systems. However, for $f$ electrons with strong spin--orbit coupling, the equation cannot yield correct wave functions. In this work, fully aware of its limitations, we nonetheless apply it to $f$ electrons.
	
	The solution of Eq.~\ref{Schrodinger} can be expressed as follows: \[\psi(r,\theta,\varphi) = R(r) Y_l^m(\theta,\varphi) \propto R(r) P_l^m(\cos\theta) e^{im\varphi},\] where $Y_l^m(\theta,\varphi)$ are spherical harmonics constructed from the associated Legendre polynomials $P_l^m(\cos\theta)$. Here, $l$ and $m$ denote the angular and magnetic quantum numbers, which also correspond to the degree and order of $P_l^m$, respectively, associated with the orbital angular momentum and its $z$-component. Throughout this paper, unless otherwise stated, $l \in \mathbb{N}$ and $m \in \mathbb{Z}$ satisfy $-l \leq m \leq l$.
	
	The $f$ electrons experience strong spin--orbit coupling, with total angular momentum given by $\mathbf{J} = \mathbf{L} + \mathbf{S}$. Consequently, the solutions of Eq.~\ref{Schrodinger} should be expressed in terms of the total angular momentum $J$ and its projection along the $z$-axis, $M$. This leads to a new function: 
	\begin{align}\label{psi_eimp}
		\psi(r,\theta,\varphi)
		&\equiv R(r) Y_J^M(\theta,\varphi) \notag \\
		&\equiv R(r) \sqrt{\frac{(2J+1)(J-M)!}{4\pi (J+M)!}} P_J^M(\cos\theta) e^{i M \varphi},
	\end{align}
	where $2J \in \mathbb{N}$ and $M \in \{-J, -J+1, \dots, J\}$, with the same conventions applied throughout. Here, $R(r)$ is a normalized radial function (see Supplemental Material, hereafter SM, for details), and the factorial is defined as $x! \equiv \Gamma(x+1)$, a convention used consistently below. We refer to the function in Eq.~\ref{psi_eimp} as the \textit{auxiliary function}. 
	
	The auxiliary function is not the true wave function of an $f$ electron, but a simplified form that captures many essential features of the actual wave function. Both share the same symmetries, and the integrals of their squared moduli are equal to one. The difference lies in continuity. While the wave function and its squared modulus are continuous, the auxiliary function becomes discontinuous when $J$ is a half-integer. Despite this, the auxiliary function is useful, as it circumvents the often cumbersome computation and allows one, through relatively simple analysis, to directly obtain the eigenstate structures of $f$ electrons under a crystal field.
	
	The squared modulus of the auxiliary function is referred to as the \textit{auxiliary density}. For $f$ electrons, its symmetry matches that of the crystal field, allowing one to infer the eigenstate structures directly from the crystal field symmetry.
	
	The $P_J^M$ in Eq.~\ref{psi_eimp} is an extension of $P_l^m$, and there exist multiple ways to generalize $P_l^m$. Legendre functions with non-integer degrees and orders have been discussed extensively \cite{bak2003half, lebedev1965special, hwang1997fully, liboff2001conical, hobson1895on, hobson1931theory, whittaker1915course, abramowitz1948handbook, Bateman1953, virchenko2001generalized, maier2016legendre, creasey2018fast, durand2022fractional, hunter1999Fermion, hunter2005properties, bildstein2018half}. The most widely used form involves hypergeometric functions \cite{lebedev1965special, hwang1997fully, liboff2001conical, hobson1895on, hobson1931theory, whittaker1915course, abramowitz1948handbook, Bateman1953, virchenko2001generalized, maier2016legendre, creasey2018fast, durand2022fractional} and was initially introduced by Hobson \cite{hobson1895on}, though this origin is often overlooked. Some works explicitly refer to these as \textit{Hobson's associated Legendre functions} \cite{whittaker1915course}. For convenience, this paper refers to them as \textit{Hobson’s Legendre functions}.
	
	For half-integer degrees and orders (as opposed to arbitrary fractional values), the generalized Rodrigues formula (SM Eq.~\ref{Rodrigues}) can be used to compute $P_J^M(x)$ \cite{bak2003half}. Since $J+M$ is a natural number, the derivative of order $J+M$ is well defined. We find that this procedure yields results identical to Hobson's Legendre functions, and thus we treat them as belonging to the same class, without discussing them separately.
	
	Hunter \textit{et al.} proposed an alternative solution \cite{hunter1999Fermion, hunter2005properties}, identifying regularities of factors for integer orders and degrees and generalizing them to half-integer cases. The resulting expressions satisfy the associated Legendre equation, and a table of factors was provided. We refer to these as \textit{Hunter's Legendre functions}.
	
	While tables are inherently limited, established patterns can lead to a general formula for arbitrary parameters. Closed-form expressions of this kind have been presented previously \cite{bildstein2018half,jepsen1955integral,An2009}, with Bildstein in particular generalizing the associated Legendre polynomials to half-integer degrees and orders \cite{bildstein2018half}. Both approaches ultimately lead to the same system, differing only in coefficients. We therefore treat Bildstein’s results as equivalent to those of Hunter \textit{et al.}.
	
	Current Legendre functions can be classified into Hobson's and Hunter's types, both satisfying the associated Legendre equation. To construct a useful auxiliary function, the Legendre function should meet four key properties (Tab.~\ref{PPP}): satisfy the associated Legendre equation, be square-integrable for normalization, exhibit similarity between functions of opposite orders, and have parity matching $J+M$ to preserve inversion symmetry in superposed states.
	
	While associated Legendre polynomials of integer degree and order satisfy all these properties, half-integer degrees and orders pose conflicts, as neither Hobson's nor Hunter's functions fully meet the requirements. Our analysis shows that introducing the sign function $\mathrm{sgn}(x)$ resolves this issue.
	
	The Legendre function in this work for half-integer $J$ and $M$ is defined from Hobson's Legendre function as follows:
	\begin{widetext}
		\begin{align}
			P_J^M(x)_{\text{this work}}
			& \equiv(-1)^{(M+|M|)[\frac{1}{2}+2J+J{\rm sgn}(x)]}\frac{(J+M)!}{(J-|M|)!}P_J^{-|M|}(x)_{\text{Hobson}} \label{tildePJM1} \\
			& =(-1)^{(M+|M|)\left[\frac{1}{2}+2J+J{\rm sgn}(x)\right]}\frac{(J+M)!}{(J+|M|)!}\frac{(1-x^2)^{\frac{|M|}{2}}}{2^J}\sum_{k=0}^{\lfloor\frac{J-|M|}{2}\rfloor}\frac{(-1)^k}{k!}\frac{(2J-2k)!}{(J-k)!}\frac{x^{J-|M|-2k}}{(J-|M|-2k)!}. \label{tildePJM2}
		\end{align}
	\end{widetext}
	
	\begin{figure}[t]
		\centering
		\includegraphics[width=3.3in]{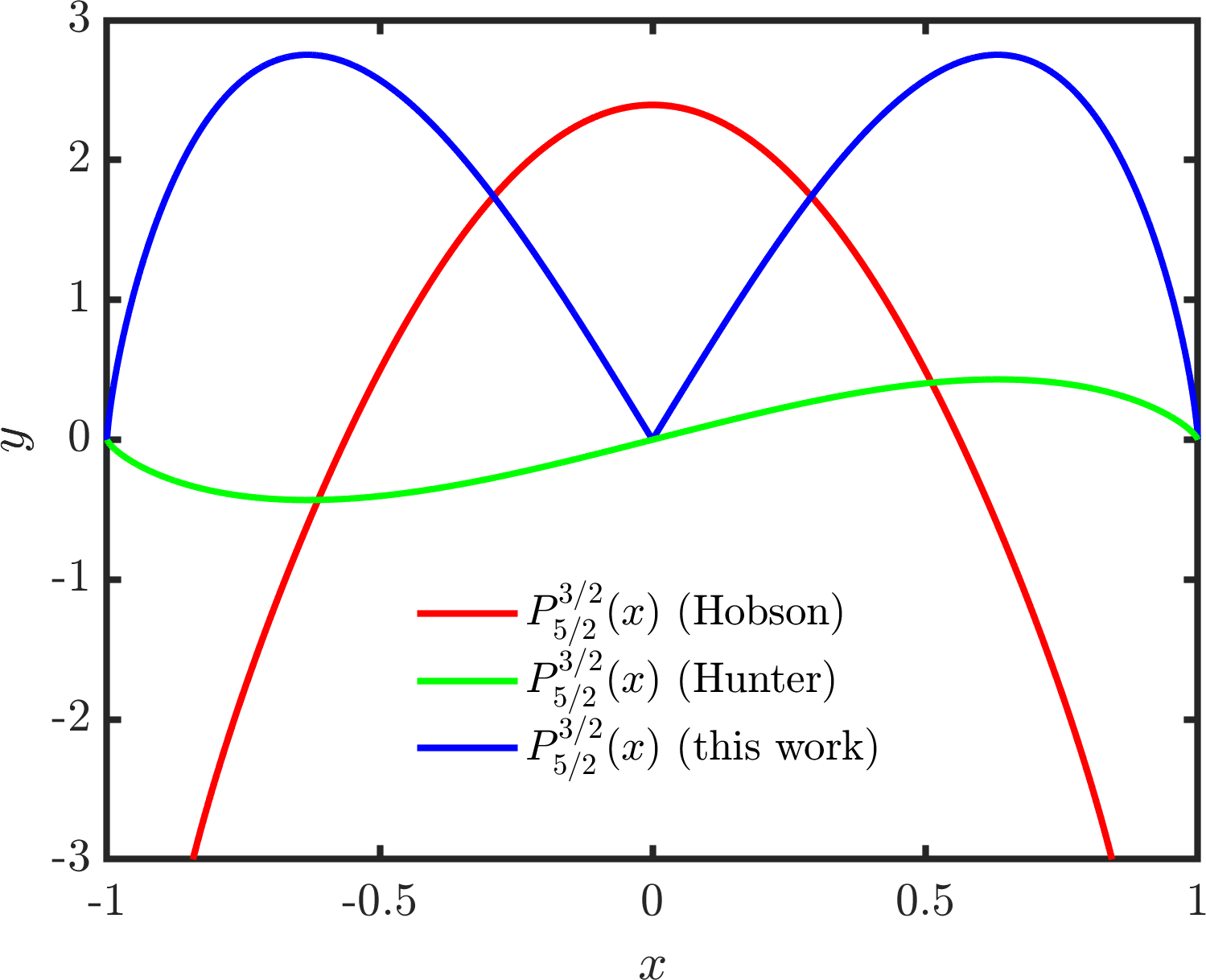}
		\caption{\label{fig:PPP}
			Comparison of $P_{5/2}^{3/2}(x)$ from Hobson \cite{hobson1895on, hobson1931theory}, Hunter \textit{et al.} \cite{hunter1999Fermion, hunter2005properties}, and the present work. Hobson's $P_{5/2}^{3/2}(x) = (8x^4 - 12x^2 + 3)/\sqrt{\frac{\pi}{2}\sqrt{(1-x^2)^3}}$ is an even function that diverges as $x \to \pm 1$, with a squared modulus that is non-integrable. Hunter's $P_{5/2}^{3/2}(x) = x (1-x^2)^{3/4}$ is an odd function and integrable. In contrast, as presented here in this work, $P_{5/2}^{3/2}(x) = 8|x|\sqrt{2/\pi}\,(1-x^2)^{3/4}$ is an even function that is integrable.
		}
	\end{figure}
	
	\begin{table}[ht]
		\centering
		\caption{Comparison of $P_J^M(x)$ for $x \neq 0$ as defined by Hobson \cite{hobson1895on, hobson1931theory}, Hunter \textit{et al.} \cite{hunter1999Fermion, hunter2005properties}, and in this work. The symbols $\checkmark$ and $\times$ denote whether the condition is always satisfied or not, respectively.
		}
		\renewcommand{\arraystretch}{1.5}
		\begin{tabular}{cccc}
			\toprule
			Property of $P_J^M(x)$ & Hobson & Hunter  & This work \\
			\midrule
			Associated Legendre equation & $\checkmark$ & $\checkmark$ & $\checkmark$ \\
			
			$\int_{-1}^1|P_J^M(x)|^2dx<\infty$ & $\times$ & $\checkmark$ & $\checkmark$ \\
			
			$\frac{\partial}{\partial x}\left|\frac{P_J^{-M}(x)}{P_J^M(x)}\right|=0$ & $\times$ & $\checkmark$ & $\checkmark$ \\
			
			$P_J^M(-x)=(-1)^{J+M}P_J^M(x)$ & $\checkmark$ & $\times$ & $\checkmark$ \\
			\bottomrule
		\end{tabular}
		\label{PPP}
	\end{table}
	
	An example is shown in Fig.~\ref{fig:PPP}. For $J = 5/2$ and $M = 3/2$, neither Hobson's nor Hunter's Legendre functions satisfy all four properties. In contrast, our function avoids the divergence present in Hobson's form while preserving the correct parity that Hunter's function lacks. Although it is singular at $x=0$, this feature is necessary to satisfy all core requirements.
	
\begin{figure}[t]
		\centering
		\includegraphics[width=3.3in]{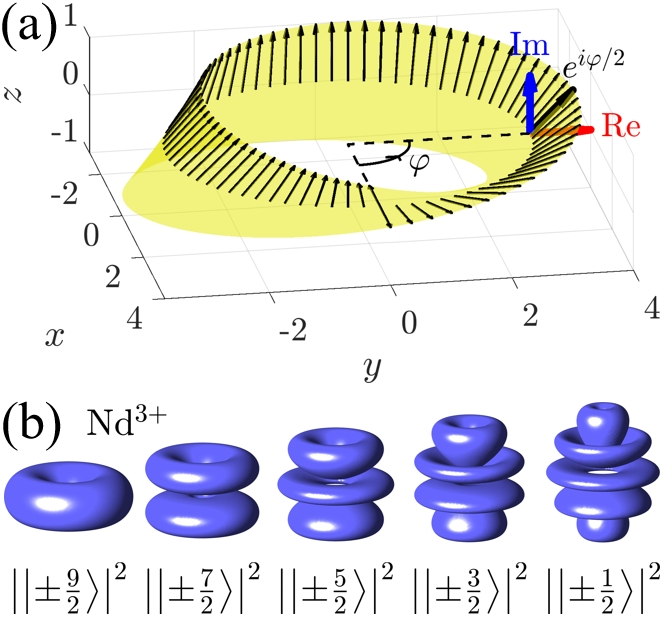}
		\caption{\label{Mobius}
			(a) The factor $e^{i\varphi/2}$ is plotted along a circular arc of radius 3, with its vector trajectory forming a M\"obius strip.  
			(b) Auxiliary densities distribution of $f$-electron states in Nd$^{3+}$ ($J=9/2$). Kets $|M\rangle$ denote the quantum states. The auxiliary densities exhibit D$_{\infty\mathrm{h}}$ symmetry, and the surface encloses 75\% of the total density.
		}
	\end{figure}

	When $J$ and $M$ are half-integers, the auxiliary function $\psi \propto P_J^M(\cos\theta)e^{iM\varphi}$ becomes antiperiodic in $\varphi$, satisfying $\psi(r,\theta,\varphi) = -\psi(r,\theta,\varphi + 2\pi)$ and thereby exhibiting a M\"obius-type topology, as illustrated in Fig.~\ref{Mobius}a. To maintain single-valuedness, continuity of the auxiliary function with respect to $\varphi$ must be relinquished. For instance, if $\varphi$ is defined on $[0, 2\pi)$, the auxiliary function is necessarily discontinuous at $\varphi = 0$. Nevertheless, the density $|\psi|^2$ remains fully periodic and continuous in $\varphi$.
	
	Hunter \textit{et al.} similarly mentioned the Möbius band when discussing half-integer spherical harmonics \cite{hunter1999Fermion}, as it arises from the $e^{iM\varphi}$ term for half-integer $M$. The difference lies in their Legendre function, whose parity does not meet the requirements of this work. 
	
	This auxiliary function approach provides an efficient and accurate tool for predicting $f$-electron eigenstates in crystal fields. An auxiliary density consistent with the local symmetry is constructed, allowing the eigenstate structures to be determined without exhaustive diagonalization. Predictions based on this framework show excellent agreement with full numerical solutions, offering both computational efficiency and insight into the electronic structure of strongly correlated $f$ electrons.
	
	\emph{Symmetry analysis.—}
	Auxiliary functions can be regarded as approximate wave functions, and one of their important characteristics is that the square modulus (auxiliary density) satisfies the crystal field symmetry. 
	Thus, auxiliary functions can be employed to determine energy eigenstates, provided their densities conform to the crystal-field symmetry. 

	Rotational symmetry is essential in constraining the eigenstates. The auxiliary density of a pure eigenstate with a single $M$ exhibits D$_{\infty\mathrm{h}}$ symmetry. A superposition, however, generally breaks this symmetry, reducing it to discrete rotational symmetries. 
	
	For instance, for a two-state superposition $\left|\psi\right\rangle = C_1\left|M_1\right\rangle + C_2\left|M_2\right\rangle$, the auxiliary density is: 
	\begin{align}
		\left|\psi\right|^2 & = \big|C_1\left|M_1\right\rangle + C_2\left|M_2\right\rangle\big|^2 \notag\\ 
		& =\left|A(r,\theta)e^{iM_1\varphi}+B(r,\theta)e^{iM_2\varphi}\right|^2 \notag\\
		& =\left(Ae^{iM_1\varphi}+Be^{iM_2\varphi}\right)\left(A^*e^{-iM_1\varphi}+B^*e^{-iM_2\varphi}\right) \notag\\
		& = |A|^2 + |B|^2 + 2|AB|\cos\left[(M_1 - M_2)\varphi + \arg(AB^*)\right], \notag
	\end{align}
	where \( A \) and \( B \) are independent of $\varphi$. The auxiliary density thus has \(|M_1 - M_2|\)-fold rotational symmetry. For example, $|M_1-M_2|=3,\,2,\,1$ give threefold, twofold, and onefold (C$_1$) symmetry, respectively.
	
	For a general superposition of \( N \) states that $\left|\psi\right\rangle = \sum_{j=1}^N A_j(r,\theta) e^{iM_j \varphi}$, the auxiliary density is: 
	\begin{align}\label{psi2j}
		&\left|\psi\right|^2
		= \left|\sum_{j=1}^N A_j(r,\theta) e^{iM_j \varphi}\right|^2 \notag \\
		&= \sum_{j=1}^N |A_j|^2 + \sum_{j \ne k} |A_j A_k| \cos\left[(M_j - M_k)\varphi + \Arg(A_j A_k^*)\right].
	\end{align}
	Therefore, the auxiliary density exhibits $n_{\mathrm{rot}}$-fold rotational symmetry if and only if the following condition holds:
	\begin{align}\label{kappa}
		n_{\mathrm{rot}} \mid \Delta M,
	\end{align}
	where $\Delta M = M_j - M_k$ for any $j \neq k$.
	From Eq.~\ref{kappa}, the eigenstate whose auxiliary density \(|\psi|^2\) exhibits \(n_{\mathrm{rot}}\)-fold rotational symmetry can be written as
	\begin{align}\label{pcjkk}
		|\psi\rangle = \sum_{j=0}^{\lfloor(2J - k)/n_{\mathrm{rot}}\rfloor} C_j\, |J - n_{\mathrm{rot}} j - k\rangle, 
	\end{align}
	where $k \in \{0,1,\dots,n_{\mathrm{rot}}-1\}$.
	
	When analyzing rotational symmetry, it should be kept in mind that the rotational symmetry order of the auxiliary density does not necessarily coincide with that of the crystal field; the former can be higher than the latter. This originates from the fact that the auxiliary density is always inversion-symmetric (see proof in SM). In particular, when the crystal field has an odd-fold rotational symmetry and a mirror plane perpendicular to the rotation axis, the rotational symmetry order of the auxiliary density becomes twice that of the crystal field. 
	As an example, for a crystal field with $\mathrm{C}_{n\mathrm{h}}$ symmetry, the symmetry of the auxiliary density falls into the following two cases:
	\begin{align}\label{ICk}
		\mathrm{C}_{\mathrm{i}} \times \mathrm{C}_{n\mathrm{h}} =
		\begin{cases}
			 \mathrm{C}_{2n\mathrm{h}}, & \text{if } n \text{ is odd}, \\
			 \mathrm{C}_{n\mathrm{h}}, & \text{if } n \text{ is even}.
		\end{cases}
	\end{align}
	Strictly speaking, $n_{\mathrm{rot}}$ in Eq.~\ref{pcjkk} denotes the rotational symmetry order of the auxiliary density.
	
	Beyond the analysis above, additional constraints can be obtained by exploiting other symmetries of the crystal field. Symmetry considerations can substantially restrict the allowed form of the eigenstates, in some cases even fixing them uniquely, depending on the specific situation. Further analysis may proceed by incorporating vertical mirror symmetries, which impose constraints on the relative phases of the coefficients appearing in the eigenstate expansion.
	
	From Eq.~\ref{psi2j}, the phase differences $\Arg(A_j A_k^*)$ determine the azimuthal positions of the density maxima. For a two-component state $\lvert\psi\rangle = C_1 \lvert M_1\rangle + C_2 \lvert M_2\rangle$, a vertical mirror plane located at $\varphi = \varphi_{\mathrm{m}}$ imposes the condition
	\begin{align}\label{tacc}
		\tan \Arg \frac{C_2}{C_1} = \tan\!\left[(M_1 - M_2)\varphi_{\mathrm{m}}\right],
	\end{align}
	where the use of the tangent removes the ambiguity associated with an overall global phase. Although the generalization to multi-component states is more involved, this example illustrates how auxiliary functions encode vertical mirror symmetry in a transparent manner.
	
	The above laws regarding the eigenstates can also be obtained using the crystal field Hamiltonian in Stevens-operator form. However, the auxiliary function provides a simpler and conceptually novel perspective.
	
	\emph{Example.—}
	We consider the optically active material NdCl$_3$ \cite{yin1999superfluorescence} as an example. The Nd$^{3+}$ ion ($J = 9/2$) is subject to a crystal field produced by nine surrounding Cl$^{-}$ ions, forming a local environment of $\mathrm{D}_{3\mathrm{h}}$ symmetry. 
	Because the auxiliary density is always inversion-symmetric, the point group $\mathrm{D}_{3\mathrm{h}}$, upon combination with $\mathrm{C}_\mathrm{i}$ in a manner analogous to Eq.~\ref{ICk}, is elevated to $\mathrm{D}_{6\mathrm{h}}$. Consequently, the rotational symmetry order is $n_{\mathrm{rot}} = 6$. From Eq.~\ref{kappa}, the eigenstates take the forms:
	\[
	C_{\pm \frac{9}{2}}|\pm \tfrac{9}{2}\rangle + C_{\mp \frac{3}{2}}|\mp \tfrac{3}{2}\rangle, \quad
	C_{\pm \frac{7}{2}}|\pm \tfrac{7}{2}\rangle + C_{\mp \frac{5}{2}}|\mp \tfrac{5}{2}\rangle, \quad
	|\pm \tfrac{1}{2}\rangle.
	\]  
	Taking into account the vertical mirror plane located at $\varphi_{\mathrm{m}} = -6.72^\circ$ (see SM Fig.~\ref{fg:NdCl3}), Eq.~\ref{tacc} yields: 
	\[
	\tan\Arg\left({C_{\pm \frac{9}{2}}}/{C_{\mp \frac{3}{2}}}\right) = \tan\Arg\left({C_{\pm \frac{7}{2}}}/{C_{\mp \frac{5}{2}}}\right) \approx \pm 0.849\ \ .
	\]  
	
	These relations, derived from the auxiliary functions, are confirmed by McPhase simulations \cite{rotter2004using}, which produce eigenstates consistent with the forms above. The extracted phase ratios,  
	\[
	\tan\Arg\left({C_{\pm \frac{9}{2}}}/{C_{\mp \frac{3}{2}}}\right) \approx \tan\Arg\left({C_{\pm \frac{7}{2}}}/{C_{\mp \frac{5}{2}}}\right) \approx \pm 0.824,
	\]  
	show excellent agreement with the predicted values. The auxiliary densities fully preserve the crystal-field symmetry, further validating the method. See SM for details.
	
	\emph{Discussion.—}
	In the presence of degeneracy, certain eigenstates may appear to deviate from the predicted pattern of the auxiliary functions, as they can form linear combinations within the degenerate subspace. A suitable basis transformation restores the expected form, as illustrated in \ce{NdCl3}, where the degenerate $\ket{\pm\tfrac{1}{2}}$ states may combine into $\tfrac{1}{\sqrt{2}}(\ket{+\tfrac{1}{2}}+\ket{-\tfrac{1}{2}})$, but agreement with the rules is recovered in the proper basis.
	
	The factor $\mathrm{sgn}(x)$ ensures the correct parity of half-integer $M$ states and thereby preserves inversion symmetry in the auxiliary density. Though for superpositions with odd rotational symmetry, it generally produces discontinuities at the equatorial plane ($\theta=90^\circ$), as in $\tfrac{1}{\sqrt{2}}(\ket{+\tfrac{1}{2}}+\ket{-\tfrac{1}{2}})$. The specific form of $\mathrm{sgn}(x)$ is not unique; any factor that enforces the correct parity for positive half-integer $M$ suffices. The expression $\bigl[\tfrac{1}{2}+2J+J\mathrm{sgn}(x)\bigr]$ in Eqs.~\eqref{tildePJM1} and \eqref{tildePJM2} could be simplified to $\bigl[\tfrac{1}{2}+J\mathrm{sgn}(x)\bigr]$, but we retain the $+2J$ term to ensure that the sign of the Legendre function near $x=0$ matches Hobson’s.
	
	Since spin $S$ and orbital angular momentum $L$ are neglected, the auxiliary functions only approximate the true $f$-electron wave functions. This approximation fails for ions such as Gd$^{3+}$, where $L=0$ and the ion forms pure states, whereas the theory predicts multi-component ones.
	
	Nevertheless, the auxiliary functions remain a practical tool, as small discontinuities or deviations in degenerate systems do not affect their usefulness for symmetry analysis and qualitative characterization of eigenstate structures.
	
	\emph{Conclusion and Outlook.—}  
	By modifying the Legendre functions, auxiliary functions with Möbius-like topologies are constructed. These functions provide an efficient framework for predicting eigenstates in $f$-electron systems. They naturally respect crystal-field symmetries, including multi-fold rotational and mirror symmetries, and offer an intuitive approach for determining eigenstate configurations. Compared with full crystal-field calculations, the auxiliary functions reproduce eigenstate structures with high accuracy, demonstrating their practical value for analyzing complex $f$-electron materials.  
	
	The predictive success of these auxiliary functions suggests they capture essential physical principles underlying $f$-electron eigenstates. While their full physical significance remains to be elucidated, this approach opens promising avenues for further exploration, including applications to other strongly correlated or geometrically frustrated systems, investigations of deeper connections with symmetry and topology, and interpretation of experimental observations in complex materials. 
	
	\begin{acknowledgments}
		\emph{Acknowledgment.—}
		The author gratefully acknowledges Hanteng Wang for valuable discussions and suggestions. The author also thanks colleagues from Southern University of Science and Technology for their assistance with the use of McPhase.
	\end{acknowledgments}
	
	\bibliography{Mobius_260125}
	
	\pagebreak
	
	\widetext
	\begin{center}
		\textbf{\large Supplemental Material for \\``M\"obius-topological auxiliary function for $f$ electrons''}
	\end{center}
	\setcounter{secnumdepth}{0}
	\setcounter{equation}{0}
	\setcounter{figure}{0}
	\setcounter{table}{0}
	\makeatletter
	\renewcommand{\theequation}{S\arabic{equation}}
	\renewcommand{\thetable}{S\arabic{table}}
	\renewcommand{\thefigure}{S\arabic{figure}}
	\renewcommand{\bibnumfmt}[1]{[S#1]}
	
	\section{radial factor of the auxiliary function}
	
	In this work, the auxiliary function $\psi(r,\theta,\varphi)$ and the spherical harmonics $Y_J^M(\theta,\varphi)$ are both normalized. Consequently, the radial function $R(r)$ in Eq.~\ref{psi_eimp} is also normalized, i.e.,
	\begin{align}
		\int_0^\infty |R(r)|^2 r^2 \, \mathrm{d}r = 1 .
	\end{align}
	
	For the Hamiltonian in Eq.~\ref{psi_eimp}, the radial function takes the form
	\begin{align}\label{RrZZ}
		R(r) = \left( \frac{2Z}{n a_0} \right)^{3/2}
		R_n^l\!\left( \frac{2Zr}{n a_0} \right),
	\end{align}
	where $n$ denotes the principal quantum number, $Z$ is the nuclear charge, and $a_0$ is the Bohr radius. The dimensionless radial function is given by
	\begin{align}
		R_n^l(x) =
		\sqrt{ \frac{(n - l - 1)!}{2n (n + l)!} }
		\, x^l e^{-x/2} L_{n - l - 1}^{2l + 1}(x),
	\end{align}
	where $L_{n - l - 1}^{2l + 1}(x)$ denotes the associated Laguerre polynomial.
	
	In this work, $Z$ is replaced by an effective nuclear charge $\tilde{Z}$, defined as
	\begin{align}
		\tilde{Z} = \frac{n}{n^*} Z^* ,
	\end{align}
	where $Z^*$ and $n^*$ are effective values determined according to Slater’s rules \cite{slater1930atomic}. This substitution accounts for electron shielding effects in multi-electron atoms, allowing the use of hydrogen-like wave functions with effective parameters.
	
	In this work, $l$ in Eq.~\ref{RrZZ} is set to 3 for the angular momentum of $f$ electrons. Whether $l$ should instead be $J$, $L$, or $S$ from spin--orbit coupling requires further study. Laguerre functions can be expressed via hypergeometric functions, but may diverge for half-integer parameters. As our analysis is insensitive to the radial distribution, we temporarily adopt $l=3$. Since the physical meaning of the auxiliary function is not fully established, the validity of this choice requires further investigation.
	
	\section{About Legendre functions}
	
	\subsection{4 important properties}
	
	In this work, we need to find or construct a Legendre function $P_J^M(x)$ that satisfies the following four properties:
	\begin{align} \label{PJM=1} 
		\left[(1-x^2)\frac{d^2}{dx^2}-2x\frac{d}{dx}+J(J+1)-\frac{M^2}{1-x^2}\right]P_J^M(x)=0,
	\end{align}
	\begin{align} \label{PJM=2} 
		\int_0^1|P_J^M(x)|^2dx=\frac{1}{2J+1}\frac{(J+M)!}{(J-M)!},
	\end{align}
	\begin{align} \label{PJM=3} 
		\left|\frac{P_J^{-M}(x)}{(J-M)!}\right|=\left|\frac{P_J^M(x)}{(J+M)!}\right|,
	\end{align}
	\begin{align} \label{PJM=4} 
		P_J^M(-x)=(-1)^{J+M}P_J^M(x).
	\end{align}
	
	These four properties hold for associated Legendre polynomials with integer degrees and orders. For half-integer degrees, constructing the corresponding functions requires additional care. Properties~(\ref{PJM=1}--\ref{PJM=4}) correspond one-to-one to those listed in Table~\ref{PPP}, with the second and third properties in the table weaker than those in Eqs.~\ref{PJM=2} and~\ref{PJM=3} to accommodate Hunter's alternative normalization convention.
	
	Eqs.~\ref{PJM=1}--\ref{PJM=4} each have specific implications. 
	Eq. \ref{PJM=1} ensures that \(Y_J^M(\theta, \varphi)\) is an eigenfunction of \(\hat{L}^2\) with eigenvalue \(J(J+1)\hbar^2\) and of \(\hat{L}_z\) with eigenvalue \(M\hbar\), namely: 
	\begin{align} 
		\hat{L}^2Y_J^M(\theta, \varphi)=-\hbar^2\left(\frac{1}{{\rm sin}\theta}\frac{\partial}{\partial \theta}{\rm sin}\theta\frac{\partial}{\partial \theta}+\frac{1}{{\rm sin}^2\theta}\frac{\partial^2}{\partial \varphi^2}\right)Y_J^M(\theta, \varphi)=J(J+1)\hbar^2Y_J^M(\theta, \varphi), \notag
	\end{align}
	\begin{align} 
		\hat{L}_zY_J^M(\theta, \varphi)=\frac{\hbar}{i}\frac{\partial}{\partial \varphi}Y_J^M(\theta, \varphi)=M\hbar Y_J^M(\theta, \varphi). \notag
	\end{align}
	
	Eq.~\ref{PJM=2} ensures that \(Y_J^M\) satisfies the normalization condition, namely: 
	\begin{align} 
		\int_0^{2\pi}\int_0^\pi|Y_J^M(\theta,\varphi)|^2{\rm sin} \theta d\theta d\varphi=1. \notag
	\end{align}
	
	Eq.~\ref{PJM=3} ensures that the auxiliary density distributions of \(\left| -M \right>\) and \(\left| +M \right>\) are identical, namely: 
	\begin{align} 
		\frac{\left|\psi_J^{-M}(r,\theta,\varphi)\right|^2}{\left|\psi_J^M(\theta,\varphi)\right|^2}=\frac{\left|Y_J^{-M}(r,\theta,\varphi)\right|^2}{\left|Y_J^M(\theta,\varphi)\right|^2}=1. \notag
	\end{align}
	
	Eq.~\ref{PJM=4} ensures that the auxiliary density of any superposition state exhibits inversion symmetry, namely: 
	\begin{align} 
		\left|\psi(r,\theta,\varphi)\right|^2=\left|\psi(\pi-\theta,\varphi+\pi)\right|^2. \notag
	\end{align}
	This is crucial for the symmetry analysis presented in this work and is proved in detail in a later section.
	
	Two important approaches exist for extending the associated Legendre polynomials to non-integer orders and degrees. The more widely used method is due to Hobson \cite{hobson1895on, hobson1931theory}, while the other was proposed by Hunter \textit{et al.} \cite{hunter1999Fermion, hunter2005properties}.
	
	\subsection{Hobson's Legendre function}
	
	\begin{table}[t]
		\caption{Hobson's $P_J^M(x)$. }
		\renewcommand{\arraystretch}{1.2}
		\centering
		\resizebox{\textwidth}{!}{
			\begin{tabular}{ccccccc}
				\toprule
				$M\downarrow$ & $J=\frac{1}{2}$ & $J=\frac{3}{2}$ & $J=\frac{5}{2}$ & $J=\frac{7}{2}$ & $J=\frac{9}{2}$ \\
				
				\midrule
				
				$\frac{9}{2}$ & & & & & $\frac{3x\left(128x^8-576x^6+1008x^4-840x^2+315\right)}{\sqrt{\frac{\pi}{2}\sqrt{\left(1-x^2\right)^9}}}$ \\
				
				$\frac{7}{2}$ & & & & $\frac{3x\left(16x^6-56x^4+70x^2-35\right)}{\sqrt{\frac{\pi}{2}\sqrt{\left(1-x^2\right)^7}}}$ & $\frac{3\left(128x^8-448x^6+560x^4-280x^2+35\right)}{\sqrt{\frac{\pi}{2}\sqrt{\left(1-x^2\right)^7}}}$ \\
				
				$\frac{5}{2}$ & & & $\frac{x\left(8x^4-20x^2+15\right)}{\sqrt{\frac{\pi}{2}\sqrt{\left(1-x^2\right)^5}}}$ & $\frac{3\left(16x^6-40x^4+30x^2-5\right)}{\sqrt{\frac{\pi}{2}\sqrt{\left(1-x^2\right)^5}}}$ & $\frac{3x\left(64x^6-168x^4+140x^2-35\right)}{\sqrt{\frac{\pi}{2}\sqrt{\left(1-x^2\right)^5}}}$ \\ 
				
				$\frac{3}{2}$ & & $\frac{x\left(2x^2-3\right)}{\sqrt{\frac{\pi}{2}\sqrt{\left(1-x^2\right)^3}}}$ & $\frac{8x^4-12x^2+3}{\sqrt{\frac{\pi}{2}\sqrt{\left(1-x^2\right)^3}}}$ & $\frac{x(24x^4-40x^2+15)}{\sqrt{\frac{\pi}{2}\sqrt{\left(1-x^2\right)^3}}}$ & $\frac{64x^6-120x^4+60x^2-5}{\sqrt{\frac{\pi}{2}\sqrt{\left(1-x^2\right)^3}}}$ \\ 
				
				$\frac{1}{2}$ & $\frac{x}{\sqrt{\frac{\pi}{2}\sqrt{1-x^2}}}$ & $\frac{2x^2-1}{\sqrt{\frac{\pi}{2}\sqrt{1-x^2}}}$ & $\frac{x\left(4x^2-3\right)}{\sqrt{\frac{\pi}{2}\sqrt{1-x^2}}}$ & $\frac{8x^4-8x^2+1}{\sqrt{\frac{\pi}{2}\sqrt{1-x^2}}}$ & $\frac{x\left(16x^4-20x^2+5\right)}{\sqrt{\frac{\pi}{2}\sqrt{1-x^2}}}$ \\
				
				
				$-\frac{1}{2}$ & $\sqrt{\frac{2}{\pi}\sqrt{1-x^2}}$ & $ x\sqrt{\frac{2}{\pi}\sqrt{1-x^2}}$ & $\frac{1}{3} \left(4x^2-1\right)\sqrt{\frac{2}{\pi}\sqrt{1-x^2}}$ & $x\left(2x^2-1\right)\sqrt{\frac{2}{\pi}\sqrt{1-x^2}}$ & $\frac{1}{5}\left(16x^4-12x^2+1\right)\sqrt{\frac{2}{\pi}\sqrt{1-x^2}}$ \\ 
				
				$-\frac{3}{2}$ & & $ \frac{1}{3}\sqrt{\frac{2}{\pi}}\left(1-x^2\right)^\frac{3}{4}$ & $\frac{x}{3}\sqrt{\frac{2}{\pi}}\left(1-x^2\right)^\frac{3}{4}$ & $\frac{1}{15}\left(6x^2-1\right)\sqrt{\frac{2}{\pi}}\left(1-x^2\right)^\frac{3}{4}$ & $\frac{x}{15}\left(8x^2-3\right)\sqrt{\frac{2}{\pi}}\left(1-x^2\right)^\frac{3}{4}$ \\ 
				
				$-\frac{5}{2}$ & & & $\frac{1}{15}\sqrt{\frac{2}{\pi}}\left(1-x^2\right)^\frac{5}{4}$ & $\frac{x}{15}\sqrt{\frac{2}{\pi}}\left(1-x^2\right)^\frac{5}{4}$ & $\frac{1}{105}\left(8x^2-1\right)\sqrt{\frac{2}{\pi}}\left(1-x^2\right)^\frac{5}{4}$ \\ 
				
				$-\frac{7}{2}$ & & & $$ & $\frac{1}{105}\sqrt{\frac{2}{\pi}} \left(1-x^2\right)^\frac{7}{4}$ & $\frac{x}{105}\sqrt{\frac{2}{\pi}} \left(1-x^2\right)^\frac{7}{4}$ \\
				
				$-\frac{9}{2}$ & & & & & $\frac{1}{945}\sqrt{\frac{2}{\pi}} \left(1-x^2\right)^\frac{9}{4}$ \\
				\bottomrule
			\end{tabular}
		}
		\label{tab:hobson}
	\end{table}
	
	Hobson's Legendre functions arise from the associated Legendre equation with complex parameters and come in two types: $P_\lambda^\mu(z)$ and $Q_\lambda^\mu(z)$. Among these, $P_\lambda^\mu(z)$ serves as the natural extension of the Legendre polynomial, with its explicit form given by \cite{hobson1895on,hobson1931theory,whittaker1915course,abramowitz1948handbook,Bateman1953,virchenko2001generalized,maier2016legendre,creasey2018fast,durand2022fractional}:
	\begin{align}
		P_\lambda^\mu(z)=\frac{1}{(-\mu)!}\left(\frac{z+1}{z-1}\right)^{\mu/2}\,_2F_1\left(-\lambda,\lambda+1;1-\mu;\frac{1-z}{2}\right), \label{Plamu}
	\end{align}
	where $\lambda$, $\mu$, and $z$ may be complex, and $_2F_1$ denotes the hypergeometric function. When $\lambda$ and $\mu$ take integer or half-integer values $J$ and $M$, $P_J^M(x)$ can be expressed as follows:
	\begin{align}\label{PJM_Hobson}
		P_J^M(x)=\frac{(J+M)!}{(J-M)!}\frac{(1-x^2)^{-M/2}}{2^J}\sum_{k=0}^{\lfloor(J+M)/2\rfloor}\frac{(-1)^k}{k!}\frac{(2J-2k)!}{(J-k)!}\frac{x^{J+M-2k}}{(J+M-2k)!}.
	\end{align}
	This result can also be derived from the generalized Rodrigues formula of the associated Legendre function, as follows: 
	\begin{align} \label{Rodrigues}
		P_J^M(x) = \frac{(-1)^M}{2^J J!} (1 - x^2)^{M/2} \dv[J + M]{x} (x^2 - 1)^J.
	\end{align}
	These expressions are consistent with Eqs.~\ref{Plamu} and \ref{PJM_Hobson} where applicable. 
	Table~\ref{tab:hobson} lists explicit forms of Hobson’s $P_J^M(x)$.
	
	The associated Legendre polynomials arise as the special case of Eq.~\ref{Plamu} with $\lambda \in \mathbb{N}$ and $\mu \in [-\lambda,\lambda]\cap\mathbb{Z}$. In this case, the prefactor $(-\lambda)$ in Eq.~\ref{Plamu} forces all terms with indices greater than $\lambda$ to vanish, truncating the infinite series to a finite sum. Consequently, the Legendre function reduces to polynomials. 
	
	Hobson's Legendre functions satisfy Eqs.~\ref{PJM=1} and \ref{PJM=4}, but for half-integer degrees and orders they fail to satisfy Eq.~\ref{PJM=3}. For half-integer orders $M\leq1$, Eq.~\ref{PJM=2} remains valid; for $M>1$, however, it breaks down and the integral diverges. Consequently, no normalization coefficient can be defined, and the functions are intrinsically non-normalizable. 
	
	Because they fail to satisfy Eqs.~\ref{PJM=2} and \ref{PJM=3}, Hobson's Legendre functions are unsuitable for the present work.
	
	\subsection{Hunter's Legendre function}
	
	Hunter \textit{et al.} proposed a formulation of Legendre function parameters valid for integer or half-integer degrees and orders \cite{hunter1999Fermion,hunter2005properties}. By analyzing the associated Legendre polynomials, they identified patterns among the factors, summarized in a table \cite{hunter1999Fermion} and reproduced in Table~\ref{HunterPJM}. These results are purely polynomial, valid only for integer or half-integer parameters, and do not extend to general fractional values. Strictly speaking, they are Hunter's Legendre polynomials, but here we refer to them as Hunter's Legendre functions for simplicity.
	
	Hunter gave $P_J^M$ only for $M \ge 0$, as negative $M$ is redundant under his convention
	\(Y_J^M \equiv P_J^{|M|}(\cos\theta) e^{i M \varphi}\)\cite{hunter2005properties},
	so one can define $P_J^{-|M|} \equiv P_J^{|M|}$. This ensures the third condition in Table~\ref{PPP}, and the integrability of $|P_J^M|^2$ satisfies the second condition. Although Hunter's functions do not exactly match Eqs.~\ref{PJM=2} and \ref{PJM=3} due to coefficient conventions, their integrability and symmetry remain valid.
	
	Hunter correctly identified the underlying patterns, and his functions do satisfy the associated Legendre equation (Eq.~\ref{PJM=1}). Therefore, Hunter's Legendre functions meet three of the four conditions, but they do not satisfy the fourth condition in Table~\ref{PPP} (Eq.~\ref{PJM=4}). If the auxiliary function is constructed using Hunter's Legendre function, the resulting auxiliary density does not necessarily preserve inversion symmetry, particularly for half-integer $J$ with mixed-sign $M$ components. Hence, Hunter's approach is not ideal, and a reformulation of the Legendre functions remains necessary.
	
	\begin{table}[t]
		\caption{Hunter's $P_J^M(x)$. $P_J^M(x)$ can be taken as $P_J^{|M|}(x)$ for $M < 0$.}
		\renewcommand{\arraystretch}{1.2}
		\centering
		\resizebox{\textwidth}{!}{
			\begin{tabular}{ccccccc}
				\toprule
				$M\downarrow$ & $J=\frac{1}{2}$ & $J=\frac{3}{2}$ & $J=\frac{5}{2}$ & $J=\frac{7}{2}$ & $J=\frac{9}{2}$ \\
				
				\midrule
				
				$\frac{9}{2}$ & & & & & $\left(1-x^2\right)^\frac{9}{4}$ \\
				
				$\frac{7}{2}$ & & & & $\left(1-x^2\right)^\frac{7}{4}$ & $x\left(1-x^2\right)^\frac{7}{4}$ \\
				
				$\frac{5}{2}$ & & & $\left(1-x^2\right)^\frac{5}{4}$ & $x\left(1-x^2\right)^\frac{5}{4}$ & $\left(1-8x^2\right)\left(1-x^2\right)^\frac{5}{4}$ \\ 
				
				$\frac{3}{2}$ & & $\left(1-x^2\right)^\frac{3}{4}$ & $x\left(1-x^2\right)^\frac{3}{4}$ & $\left(1-6x^2\right)\left(1-x^2\right)^\frac{3}{4}$ & $x\left(3-8x^2\right)\left(1-x^2\right)^\frac{3}{4}$ \\ 
				
				$\frac{1}{2}$ & $\sqrt[4]{1-x^2}$ & $ x\sqrt[4]{1-x^2}$ & $\left(1-4x^2\right)\sqrt[4]{1-x^2}$ & $3x\left(1-2x^2\right)\sqrt[4]{1-x^2}$ & $3\left(16x^4-12x^2+1\right)\sqrt[4]{1-x^2}$ \\
				\bottomrule
			\end{tabular}
		}
		\label{HunterPJM}
	\end{table}
	
	\subsection{Legendre function in this work}
	
	Existing Legendre functions do not satisfy all four conditions, Eqs.~\ref{PJM=1}--\ref{PJM=4}. 
	For half-integer \(J\) and \(M\), Eqs.~\ref{PJM=3} and \ref{PJM=4} conflict; 
	for example, when \(M=\pm 1/2\), one requires opposite parity while the other requires the same. 
	This is resolved by introducing the factor \(\operatorname{sgn}(x)\).
	
	In this work, the Legendre function is defined as
	\begin{align}\label{tildePJM}
		P_\lambda^\mu(x)_{\textrm{this work}}
		\equiv (-1)^{(\mu+|\mu|)\left[\frac{1}{2}+2\lambda+\lambda\,{\rm sgn}(x)\right]}
		\frac{(\lambda+\mu)!}{(\lambda-|\mu|)!} P_\lambda^{-|\mu|}(x)_{\rm Hobson},
	\end{align}
	where \(\lambda\) and \(\mu\) may be complex.
	
	For half-integer $\lambda$ and $\mu$ corresponding to $J$ and $M$, Eq.~\ref{tildePJM} reduces to Eq.~\ref{tildePJM1}, yielding Eq.~\ref{tildePJM2} via Eq.~\ref{PJM_Hobson}.
	Table~\ref{PJMthiswork} lists examples of \(P_J^M(x)\) for \(x\neq 0\). 
	The Legendre functions defined here satisfy all four conditions, with occasional discontinuity at \(x=0\).
	
	Introducing $\operatorname{sgn}(x)$ flips the parity for positive half-integer $M$, sometimes causing a discontinuity at $x=0$. 
	The exponent of $(-1)$ in Eq.~\ref{tildePJM} is taken as 
	$\frac{1}{2}+2\lambda+\lambda\,\operatorname{sgn}(x)$ to match the sign of Hobson's Legendre function near $x=0$:
	\[
	\lim_{\delta\to 0} \operatorname{sgn}\!\left(
	\frac{P_J^M(\pm\delta)_{\rm This\ work}}{P_J^M(\pm\delta)_{\rm Hobson}}
	\right)=1.
	\]
	
	The explicit $\operatorname{sgn}(x)$ factor, absent in prior literature, makes our Legendre function a fundamentally new and practical tool. It applies to both half-integer and integer cases and provides valuable insights into symmetry properties.
	
	\begin{table}[t]
		\caption{$P_J^M(x)$ in this work for $x=\pm|x|$ $(x\neq 0)$.}
		\renewcommand{\arraystretch}{1.2}
		\centering
		\resizebox{\textwidth}{!}{
			\begin{tabular}{ccccccc}
				\toprule
				$M\downarrow$ & $J=\frac{1}{2}$ & $J=\frac{3}{2}$ & $J=\frac{5}{2}$ & $J=\frac{7}{2}$ & $J=\frac{9}{2}$ \\
				
				\midrule
				
				$\frac{9}{2}$ & & & & & $\pm 384\sqrt{\frac{2}{\pi}} \left(1-x^2\right)^\frac{9}{4}$ \\
				
				$\frac{7}{2}$ & & & & $\mp 48\sqrt{\frac{2}{\pi}} \left(1-x^2\right)^\frac{7}{4}$ & $\pm 384x\sqrt{\frac{2}{\pi}} \left(1-x^2\right)^\frac{7}{4}$ \\
				
				$\frac{5}{2}$ & & & $\pm 8\sqrt{\frac{2}{\pi}}\left(1-x^2\right)^\frac{5}{4}$ & $\mp 48x\sqrt{\frac{2}{\pi}}\left(1-x^2\right)^\frac{5}{4}$ & $\pm 24\left(8x^2-1\right)\sqrt{\frac{2}{\pi}}\left(1-x^2\right)^\frac{5}{4}$ \\ 
				
				$\frac{3}{2}$ & & $\mp 2\sqrt{\frac{2}{\pi}}\left(1-x^2\right)^\frac{3}{4}$ & $\pm 8x\sqrt{\frac{2}{\pi}}\left(1-x^2\right)^\frac{3}{4}$ & $\mp 4\left(6x^2-1\right)\sqrt{\frac{2}{\pi}}\left(1-x^2\right)^\frac{3}{4}$ & $\pm 8x\left(8x^2-3\right)\sqrt{\frac{2}{\pi}}\left(1-x^2\right)^\frac{3}{4}$ \\ 
				
				$\frac{1}{2}$ & $\pm\sqrt{\frac{2}{\pi}\sqrt{1-x^2}}$ & $\mp 2x\sqrt{\frac{2}{\pi}\sqrt{1-x^2}}$ & $\pm \left(4x^2-1\right)\sqrt{\frac{2}{\pi}\sqrt{1-x^2}}$ & $\mp 4x\left(2x^2-1\right)\sqrt{\frac{2}{\pi}\sqrt{1-x^2}}$ & $\pm \left(16x^4-12x^2+1\right)\sqrt{\frac{2}{\pi}\sqrt{1-x^2}}$ \\
				
				$-\frac{1}{2}$ & $\sqrt{\frac{2}{\pi}\sqrt{1-x^2}}$ & $ x\sqrt{\frac{2}{\pi}\sqrt{1-x^2}}$ & $\frac{1}{3} \left(4x^2-1\right)\sqrt{\frac{2}{\pi}\sqrt{1-x^2}}$ & $x\left(2x^2-1\right)\sqrt{\frac{2}{\pi}\sqrt{1-x^2}}$ & $\frac{1}{5}\left(16x^4-12x^2+1\right)\sqrt{\frac{2}{\pi}\sqrt{1-x^2}}$ \\ 
				
				$-\frac{3}{2}$ & & $ \frac{1}{3}\sqrt{\frac{2}{\pi}}\left(1-x^2\right)^\frac{3}{4}$ & $\frac{x}{3}\sqrt{\frac{2}{\pi}}\left(1-x^2\right)^\frac{3}{4}$ & $\frac{1}{15}\left(6x^2-1\right)\sqrt{\frac{2}{\pi}}\left(1-x^2\right)^\frac{3}{4}$ & $\frac{x}{15}\left(8x^2-3\right)\sqrt{\frac{2}{\pi}}\left(1-x^2\right)^\frac{3}{4}$ \\ 
				
				$-\frac{5}{2}$ & & & $\frac{1}{15}\sqrt{\frac{2}{\pi}}\left(1-x^2\right)^\frac{5}{4}$ & $\frac{x}{15}\sqrt{\frac{2}{\pi}}\left(1-x^2\right)^\frac{5}{4}$ & $\frac{1}{105}\left(8x^2-1\right)\sqrt{\frac{2}{\pi}}\left(1-x^2\right)^\frac{5}{4}$ \\ 
				
				$-\frac{7}{2}$ & & & $$ & $\frac{1}{105}\sqrt{\frac{2}{\pi}} \left(1-x^2\right)^\frac{7}{4}$ & $\frac{x}{105}\sqrt{\frac{2}{\pi}} \left(1-x^2\right)^\frac{7}{4}$ \\
				
				$-\frac{9}{2}$ & & & & & $\frac{1}{945}\sqrt{\frac{2}{\pi}} \left(1-x^2\right)^\frac{9}{4}$ \\
				\bottomrule
			\end{tabular}
		}
		\label{PJMthiswork}
	\end{table}
	
	\section{Proof of inversion symmetry in the auxiliary density}
	
	The auxiliary density $|\psi|^2$ of a superposition is always inversion-symmetric.
	Eq.~\ref{PJM=4} plays a crucial role in this property. 
	We will use it to prove the inversion symmetry of the auxiliary density in this section.
	
	For a pure state, the auxiliary density is: 
	\begin{align}\label{eq:pAPe_1}
		\left|\psi(r,\theta,\varphi)\right|^2=\left|A(r)P_J^{M}({\rm cos}\theta)e^{iM\varphi}\right|^2
	\end{align}
	For cases in which $J$ and $M$ are both integers or half-integers (as below), with Eq.~\ref{PJM=4}, the auxiliary density at the inversion position can be derived using Eq.~\ref{PJM=4} as follows: 
	\begin{align}\label{eq:pAPe_2}
		\left|\psi(r,\pi-\theta,\varphi+\pi)\right|^2=\left|A(r)P_J^{M}(-{\rm cos}\theta)\left(-e^{iM\varphi}\right)\right|^2=\left|A(r)P_J^{M}({\rm cos}\theta)e^{iM\varphi}\right|^2=\left|\psi(r,\theta,\varphi)\right|^2. 
	\end{align}
	Eq.~\ref{eq:pAPe_2} reproduces the result of Eq.~\ref{eq:pAPe_1}, thereby confirming the inversion symmetry of the auxiliary density for a pure state.
	
	Consider the superposition of two states $\left|\psi\right>=|C_1\left|M_1\right>+C_2\left|M_2\right>$, the auxiliary density is: 
	\begin{align}
		&\left|\psi(r,\theta,\varphi)\right|^2\notag \\
		=&\left|C_1\left|M_1\right>+C_2\left|M_2\right>\right|^2 \notag \\
		=&\left|A(r)P_J^{M_1}({\rm cos}\theta)e^{iM_1\varphi}+B(r)P_J^{M_2}({\rm cos}\theta)e^{iM_2\varphi}\right|^2 \notag \\
		=&\left|AP_J^{M_1}({\rm cos}\theta)\right|^2+\left|BP_J^{M_2}({\rm cos}\theta)\right|^2 \notag \\
		&+2\left|ABP_J^{M_1}({\rm cos}\theta)P_J^{M_2}({\rm cos}\theta)\right|
		{\rm cos}\left\{(M_1-M_2)\varphi +{\rm Arg}\left[AP_J^{M_1}({\rm cos}\theta)B^*P_J^{M_2}({\rm cos}\theta)^*\right]\right\} \notag \\
		=&\left|AP_J^{M_1}({\rm cos}\theta)\right|^2+\left|BP_J^{M_2}({\rm cos}\theta)\right|^2 \notag \\
		&+2\left|ABP_J^{M_1}({\rm cos}\theta)P_J^{M_2}({\rm cos}\theta)\right|
		{\rm cos}\left\{(M_1-M_2)\varphi +{\rm Arg}\left(AB^*\right)
		+{\rm Arg}\left[P_J^{M_1}({\rm cos}\theta)P_J^{M_2}({\rm cos}\theta)\right]\right\}. \label{rho_tp_1}
	\end{align}
	Here, $A$ and $B$ depend only on $r$ and are independent of $\theta$ and $\varphi$. The auxiliary density at the inversion position is: 
	\begin{align}
		&\left|\psi(r,\pi-\theta,\varphi+\pi)\right|^2 \notag \\
		=&\left|AP_J^{M_1}(-{\rm cos}\theta)\right|^2+\left|BP_J^{M_2}(-{\rm cos}\theta)\right|^2 \notag \\
		&+2\left|ABP_J^{M_1}(-{\rm cos}\theta)P_J^{M_2}(-{\rm cos}\theta)\right|
		{\rm cos}\left\{(M_1-M_2)\left(\varphi+\pi\right) +{\rm Arg}\left(AB^*\right)
		+{\rm Arg}\left[P_J^{M_1}(-{\rm cos}\theta)P_J^{M_2}(-{\rm cos}\theta)\right]\right\} \label{rho_tp_2b} \\
		=&\left|AP_J^{M_1}({\rm cos}\theta)\right|^2+\left|BP_J^{M_2}({\rm cos}\theta)\right|^2 \notag \\
		&+2\left|ABP_J^{M_1}({\rm cos}\theta)P_J^{M_2}({\rm cos}\theta)\right|
		{\rm cos}\left\{(M_1-M_2)\left(\varphi+\pi\right) +{\rm Arg}\left(AB^*\right)
		+{\rm Arg}\left[(-1)^{J+M_1}P_J^{M_1}({\rm cos}\theta)\cdot(-1)^{J+M_2}P_J^{M_2}({\rm cos}\theta)\right]\right\} \label{rho_tp_2c} \\
		=&\left|AP_J^{M_1}({\rm cos}\theta)\right|^2+\left|BP_J^{M_2}({\rm cos}\theta)\right|^2+2\left|ABP_J^{M_1}({\rm cos}\theta)P_J^{M_2}({\rm cos}\theta)\right|\notag\\
		&\qquad\qquad\qquad\qquad\cdot{\rm cos}\left\{(M_1-M_2)\varphi +{\rm Arg}\left(AB^*\right)
		+{\rm Arg}\left[P_J^{M_1}({\rm cos}\theta)P_J^{M_2}({\rm cos}\theta)\right]
		+(M_1-M_2)\pi+{\rm Arg}\left[(-1)^{2J+M_1+M_2}\right]\right\} \notag \\
		=&\left|AP_J^{M_1}({\rm cos}\theta)\right|^2+\left|BP_J^{M_2}({\rm cos}\theta)\right|^2+2\left|ABP_J^{M_1}({\rm cos}\theta)P_J^{M_2}({\rm cos}\theta)\right|\notag\\
		&\qquad\qquad\qquad\qquad\cdot{\rm cos}\left\{(M_1-M_2)\varphi +{\rm Arg}\left(AB^*\right)
		+{\rm Arg}\left[P_J^{M_1}({\rm cos}\theta)P_J^{M_2}({\rm cos}\theta)\right]
		+(M_1-M_2)\pi+(2J+M_1+M_2)\pi\right\} \notag \\
		=&\left|AP_J^{M_1}({\rm cos}\theta)\right|^2+\left|BP_J^{M_2}({\rm cos}\theta)\right|^2\notag\\
		&+2\left|ABP_J^{M_1}({\rm cos}\theta)P_J^{M_2}({\rm cos}\theta)\right|
		{\rm cos}\left\{(M_1-M_2)\varphi +{\rm Arg}\left(AB^*\right)
		+{\rm Arg}\left[P_J^{M_1}({\rm cos}\theta)P_J^{M_2}({\rm cos}\theta)\right]
		+2\pi(J+M_1)\right\} \notag \\
		=&\left|AP_J^{M_1}({\rm cos}\theta)\right|^2+\left|BP_J^{M_2}({\rm cos}\theta)\right|^2\notag\\
		&+2\left|ABP_J^{M_1}({\rm cos}\theta)P_J^{M_2}({\rm cos}\theta)\right|
		{\rm cos}\left\{(M_1-M_2)\varphi +{\rm Arg}\left(AB^*\right)
		+{\rm Arg}\left[P_J^{M_1}({\rm cos}\theta)P_J^{M_2}({\rm cos}\theta)\right]\right\} \label{rho_tp_2} \\
		=&\left|\psi(r,\theta,\varphi)\right|^2. \notag
	\end{align}
	
	Eq.~\ref{rho_tp_2} reproduces the result of Eq.~\ref{rho_tp_1}, completing the proof. The transition from Eq.~\ref{rho_tp_2b} to Eq.~\ref{rho_tp_2c} relies directly on Eq.~\ref{PJM=4}. For superpositions of more than two states, the proof proceeds analogously and is omitted.
	
	Strictly speaking, Eq.~\ref{PJM=4} is not a necessary and sufficient condition. A weaker requirement that still ensures inversion symmetry of a superposition is:  
	\begin{align}\label{PJM=4low}
		\frac{P_J^{M+1}(-x)}{P_J^{M+1}(x)}=-\frac{P_J^M(-x)}{P_J^M(x)}.
	\end{align}
	Hunter’s Legendre functions do not satisfy Eq.~\ref{PJM=4low} for $M=-\tfrac{1}{2}$. With these functions, the corresponding density retains inversion symmetry only when all $M$ values in the superposition share the same sign. Hence, Hunter’s functions are not applicable in this context, and our conclusions remain unaffected.  
	
	We have thus demonstrated that, for both pure states and superpositions, and for angular momenta that are either integer or half-integer, the auxiliary density universally exhibits inversion symmetry:
	\begin{align}
		\left|\psi(r,\theta,\varphi)\right|^2 = \left|\psi(r,\pi-\theta,\varphi+\pi)\right|^2. \notag
	\end{align}

\section{Detailed example of \ce{NdCl3}}

We provide additional details underlying the example discussed in the main text.

Starting from the inversion symmetry of the auxiliary density, the combination
$\mathrm{C}_\mathrm{i} \times \mathrm{D}_{3\mathrm{h}} = \mathrm{D}_{6\mathrm{h}}$
implies a sixfold rotational symmetry ($n_{\mathrm{rot}}=6$). Substituting
$J=9/2$ into Eq.~\ref{pcjkk} yields the symmetry-allowed eigenstate structures:
\begin{align}\label{eq:NdCl3}
	\lvert \psi \rangle =
	\begin{cases}
		C_{\pm \frac{9}{2}} \lvert \pm \frac{9}{2} \rangle + C_{\mp \frac{3}{2}} \lvert \mp \frac{3}{2} \rangle, \\[2pt]
		C_{\pm \frac{7}{2}} \lvert \pm \frac{7}{2} \rangle + C_{\mp \frac{5}{2}} \lvert \mp \frac{5}{2} \rangle, \\[2pt]
		\lvert \pm \frac{1}{2} \rangle.
	\end{cases}
\end{align}
The states $\lvert \pm \tfrac{1}{2} \rangle$ remain isolated, as their magnetic quantum numbers differ by less than six from all other basis states.

\begin{figure}[t]
	\centering
	\includegraphics[width=5in]{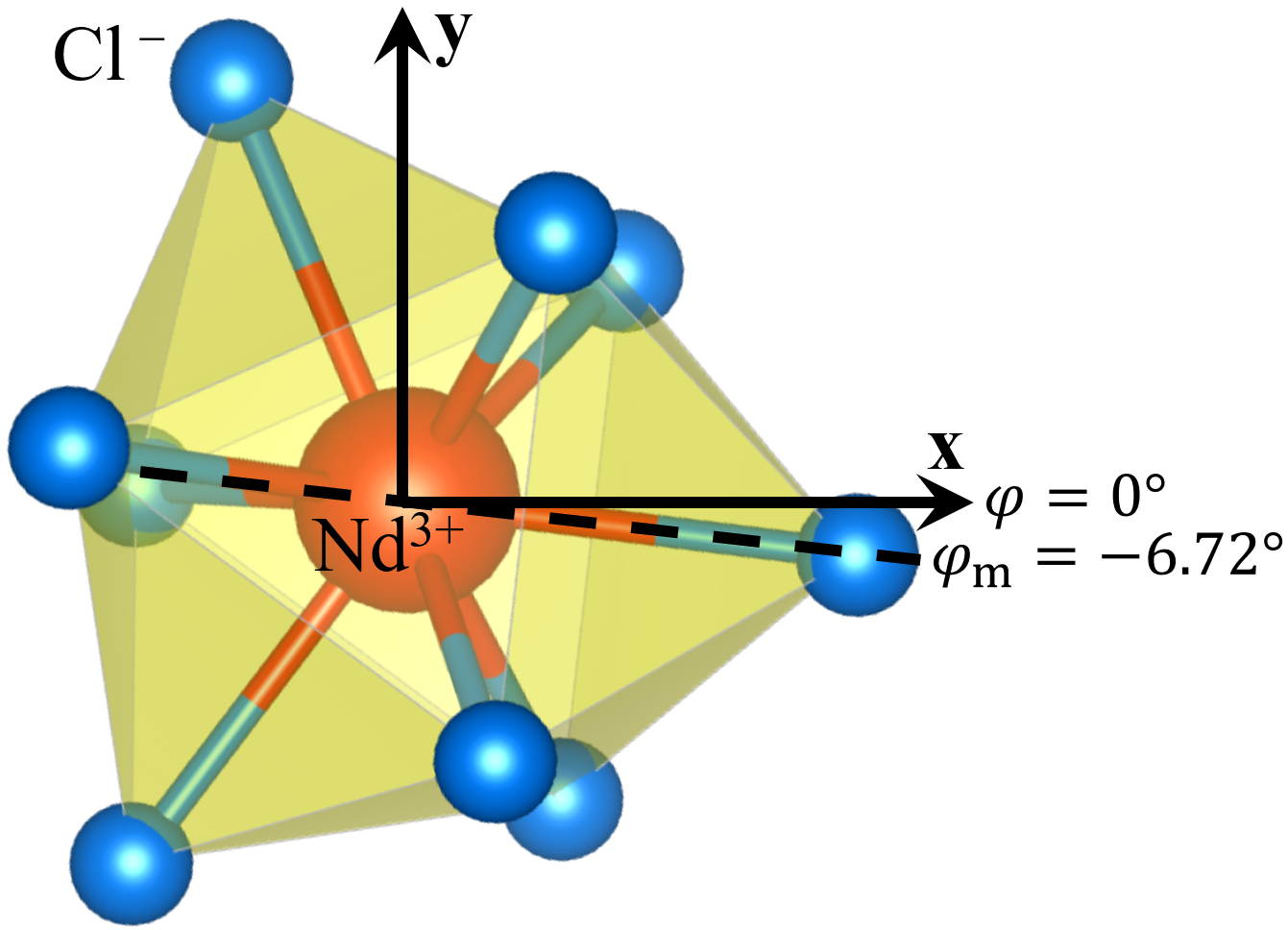}
	\caption{\label{fg:NdCl3}
		Ionic structure around Nd$^{3+}$ in NdCl$_3$, showing one of the vertical mirrors at $\varphi_{\rm m} = -6.72^\circ$.
	}
\end{figure}

Imposing the vertical mirror symmetry at $\varphi_{\mathrm{m}}=-6.72^\circ$ (Fig.~\ref{fg:NdCl3}) and using Eq.~\ref{tacc}, one obtains
\begin{align}\label{tacctacc}
	\text{tan Arg}\frac{C_{\pm\frac{9}{2}}}{C_{\mp\frac{3}{2}}}=\text{tan Arg}\frac{C_{\pm\frac{7}{2}}}{C_{\mp\frac{5}{2}}}=\mp {\rm tan}(6\varphi_{\rm m})\approx\pm 0.849\quad.
\end{align}

These results are further validated by McPhase, a software package for computing static and dynamic magnetic properties of rare-earth compounds \cite{rotter2004using}. With McPhase, the precise ground and excited states of NdCl$_3$ can be obtained as follows: 
\begin{align}\label{eq:01234}
	\begin{cases}
		\left|\psi_0\right>_\pm=0.9989\left|\pm\frac{9}{2}\right>+(0.0359\mp0.0296i)\left|\mp\frac{3}{2}\right>\\
		\left|\psi_1\right>_\pm=0.9909\left|\pm\frac{7}{2}\right>+(0.1037\mp0.0855i)\left|\mp\frac{5}{2}\right>\\
		\left|\psi_2\right>_\pm=(-0.1037\mp0.0855i)\left|\pm\frac{7}{2}\right>+0.9909\left|\mp\frac{5}{2}\right>\\
		\left|\psi_3\right>_\pm=(-0.0359\mp0.0296i)\left|\pm\frac{9}{2}\right>+0.9989\left|\mp\frac{3}{2}\right>\\
		\left|\psi_4\right>_\pm=\left|\pm\frac{1}{2}\right>
	\end{cases}. 
\end{align}
All eigenstates conform to Eq.~\ref{eq:NdCl3}, confirming the validity of the auxiliary-function result.

	The coefficients in Eq.~\ref{eq:01234} can be calculated as follows: 
\begin{align}\label{tacctacc2}
	\text{tan Arg}\frac{C_{\pm\frac{9}{2}}}{C_{\mp\frac{3}{2}}}=\text{tan Arg}\frac{C_{\pm\frac{7}{2}}}{C_{\mp\frac{5}{2}}}\approx\pm 0.824\quad.
\end{align}
A comparison between Eqs.~\ref{tacctacc} and \ref{tacctacc2} shows that the prediction in Eq.~\ref{tacctacc} is highly accurate, exhibiting only a very small discrepancy from Eq.~\ref{tacctacc2}, likely due to the approximations in the McPhase computation.

\begin{figure}[t]
	\centering
	\includegraphics[width=6in]{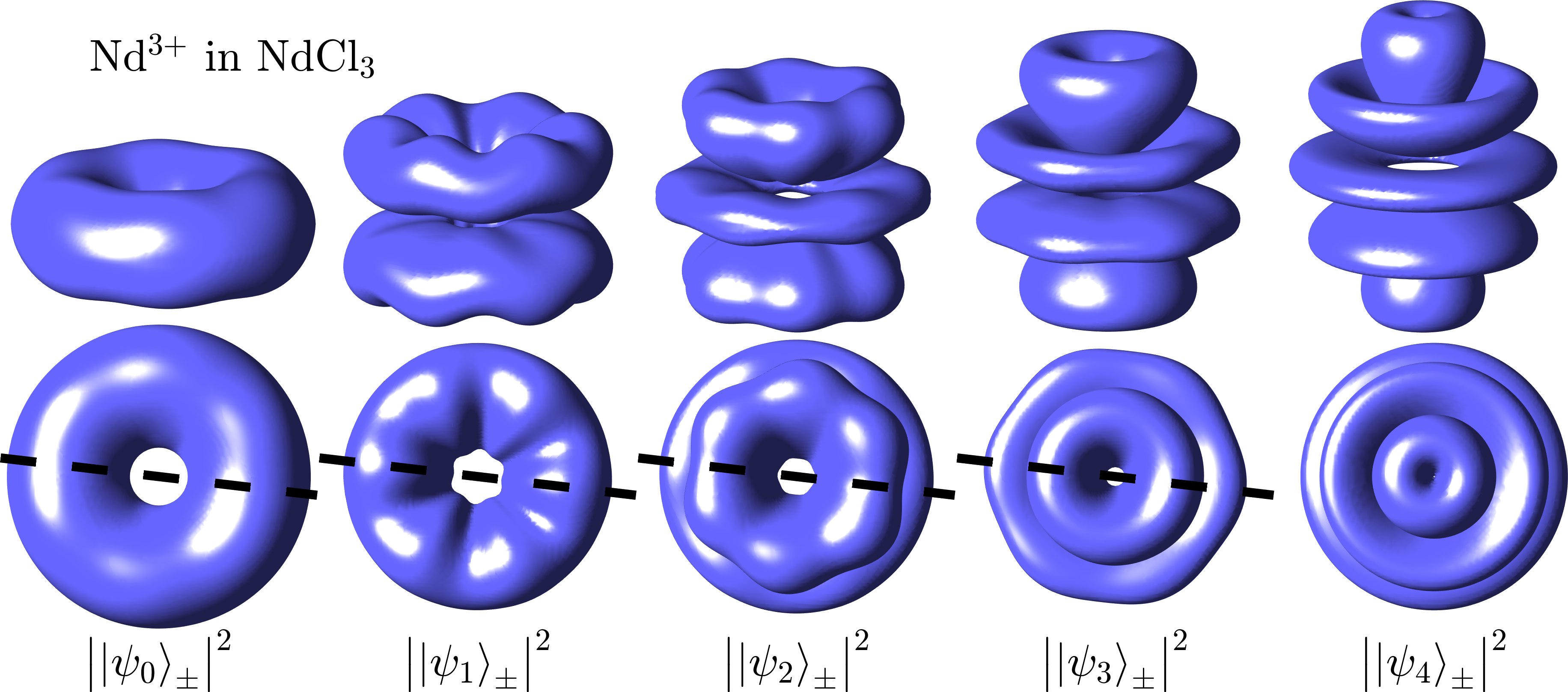}
	\caption{\label{fg:NdCl3b}
		Auxiliary densities of the eigenstates of NdCl$_3$. They all satisfy the symmetries of the crystal field. 
		The dashed line in each relevant subplot indicates one of the vertical mirrors, consistent with Fig.~\ref{fg:NdCl3}. $\bigl||\psi_4\bigr>_\pm\big|^2$ has vertical mirror symmetry at any angle. 
		The surface represents the equiprobability density surface that encloses 75\% of the total auxiliary density. 
	}
\end{figure}

We have plotted the auxiliary density distribution corresponding to the eigenstates of NdCl$_3$, as shown in Fig.~\ref{fg:NdCl3b}. The distribution fully respects the symmetry of the crystal field.

\end{document}